\documentclass[final,3p,times,twocolumn]{elsarticle}

\usepackage{graphicx}

\usepackage{amssymb}

\journal{Physics Letters B}

\newcommand{\be}{\begin{equation}}
\newcommand{\ee}[1]{\label{#1} \end{equation}}
\newcommand{\ba}{\begin{eqnarray}}
\newcommand{\ea}[1]{\label{#1} \end{eqnarray}}
\newcommand{\nl}{\nonumber \\}

\newcommand{\dd}{\textrm{d \,}}

\begin{document}

\begin{frontmatter}


 \ead{karoly.uermoessy@cern.ch}

\title{Generalised Tsallis Statistics in Electron-Positron Collisions}

 \author{K. Urmossy$^{a,b}$, G. G. Barnaf\"oldi$^b$ and T.~S.~Bir\'o$^b$}
 \address[label1]{E\"otv\"os University, \\ 1/A P\'azm\'any P\'eter s\'et\'any, H-1117 Budapest, Hungary}
 \address[label2]{KFKI Research Institute for Particle and Nuclear Physics of the HAS, \\ 
29-33 Konkoly--Thege Mikl\'os Str., H-1121 Budapest, Hungary}




\begin{abstract}
The scaling of charged hadron fragmentation functions to the Tsallis distribution for the momentum fraction $0.01 \lessapprox x \lessapprox 0.2$ is presented for various $e^+e^-$ collision energies. A possible microcanonical generalisation of the Tsallis distribution is proposed, which gives good agreement with measured data up to $x\approx1$. The proposal is based on superstatistics and a Koba\,--\,Nielsen\,--\,Olesen (KNO) like scaling of multiplicity distributions in $e^+e^-$ experiments.

\end{abstract}


\end{frontmatter}



\date{\today}

\section{Introduction}
\label{sec:int}



The question: 'why do thermal models work in high energy collisions' may be answered on a purely mathematical basis. Thermal models are based on the canonical distribution e.g. the Boltzmann-Gibbs (BG) (or generalisations of it \cite{bib:next1,bib:tsinppNN10,BiroVan:2011}). The canonical BG distributions follows directly from the \textit{Large Deviations Theorem} (LDT) \cite{bib:math1,bib:math2,bib:uk1} if the ensemble consists of a large number of independent and identically distributed particles with fixed total energy. This theorem, however, does not deal with the reason behind the identical distribution of these particles: whether it is due to the fact that they are \textit{thermal particles}, or that they are an ensemble of \textit{identically produced particles}. In any case, their total energy is conserved, either during their production, or during their thermalisation. This is probably the reason why Hagedorn's model on thermal hadron production \cite{bib:BGinppNN1} does not only work for heavy-ion collisions (for a few examples see Refs.~\cite{bib:BGinppNN1}-\cite{bib:BGinee}, and~\cite{bib:tsinppNN5}), but for $e^+e^-$ annihilations as well, which are conjectured to be non-equilibrium phenomena.


However, the BG statistics describes uncorrelated and non-interacting particles, and neither of these features are characteristics of particles stemming from high energy collisions. Apparently, the BG distribution fails to describe transverse momentum ($p_T$) spectra for intermediate and high momenta ($p_T \gtrapprox 3$ GeV/c), while the cut power-law Tsallis distribution (TS) gives a good fit in a wide measured range of proton-proton ($pp$), proton-antiproton ($p\bar{p}$), nucleus-nucleus ($AA$)~\cite{bib:tsinppNN1}-\cite{bib:tsinppNN10}, and $e^+e^-$ \cite{bib:tsinee1,bib:tsinee2} collisions. Furthermore, charged hadron fragmentation functions also scale to the TS distribution for $0.01 \lessapprox x \lessapprox 0.2$, as shown in Section~\ref{sec:res}, where $x$ stands for the momentum fraction of the produced hadron, $x = 2 p/\sqrt{s}$, with $\sqrt{s}$ being the center of mass energy.

Though numerous proposed explanations exist for the emergence of TS statistics in equilibrium and out of equilibrium systems (for reviews see Refs.~\cite{bib:tsinppNN10,bib:next1}), the reason for its occurence in high energy physics has not yet been clarified. Moreover, the TS distribution, being a generalisation of the canonical BG, has a non-compact support, and thus cannot be valid close to $x\approx1$, where it should have a cut due to the finiteness of the collision energy. Therefore, in order to achieve good agreement with data for large $x$, microcanonical effects, due to the finite available phase space, have to be taken into account, as it has also been proposed in \cite{bib:micro1}-\cite{bib:micro6}. In this paper, a possible microcanonical generalisation of the TS distribution is proposed, which gives a good fit to data on fragmentation functions measured in $e^+e^-$ collisions for $0.01 \lessapprox x \lessapprox1$.

This paper is organised as follows: in Section~\ref{sec:calc} it is shown that multiplicity distributions, similar to the \textit{Koba\,--\,Nielsen\,--\,Olesen (KNO)} scaling ones reported in Ref.~\cite{bib:kno}, can lead to a momentum distribution of the canonical TS type, or to a generalisation of it, if the momentum distribution is BG or microcanonical for a fixed multiplicity. Section~\ref{sec:res} contains fits of multiplicity distributions and fragmentation functions, used and obtained in Section~\ref{sec:calc}, to $e^+e^-$ data measured at various collision energies. Section~\ref{sec:con} contains the concluding remarks.

\section{Canonical and microcanonical Tsallis statistics from multiplicity fluctuations}
\label{sec:calc}
The fragmentation function measured in $e^+e^-$ experiments is the average of charged hadron yields produced in millions of events. For collision energies $\sqrt{s}\geq M_Z$ ($M_Z$ is the mass of the Z boson), more than  90 \% of the events are 2-jet events (a $q$ and an $\bar{q}$ jet with the same energy, $E=\sqrt{s}/2$). Hadrons in a $q$ jet have a very narrow distribution arround the jet axis in momentum space, therefore they may be considered as a one-dimensional ensemble. Neglecting the mass of the produced hadrons, energy and momentum conservation in one dimension has the same form: $\sum \epsilon_i = E$ (note that all hadrons go in the same direction, and $\epsilon_i = |\vec{p_i}|$). If we prescribe nothing else than energy-momentum conservation during the hadronisation, the produced particles become a microcanonical ensemble in one dimension. In this section we show that for massless particles,

\begin{description}
\item[a)] if the momentum distribution in events with fixed multiplicity is BG, and the multiplicity has Gamma-distribution, the average momentum distribution is TS-like;

\item[b)] if the momentum distribution in events with fixed multiplicity is microcanonical, and the shifted multiplicity ($N-N_0$) has Gamma-distribution, the approximate average momentum distribution can be considered as a possible microcanonical generalisation of the TS distribution.
\end{description}

In case, \textit{\textbf{a}}, the normalised one-particle distribution in events with multiplicity $N$, is
\be
 f_N(\epsilon) = A_c\; e^{-\beta_N \epsilon} \ \ ,
\ee{calc1}
where $A_c = \beta^D_N / \left( \, k_D\, \Gamma(D)\, \right) $ and $k_D = \int d\Omega_p$ follow from the conditions
\ba
1 &=& \int d\Omega_p \int dp\,p^{D-1} f_N(\epsilon),\nl 
\frac{E}{N} &=& \int d\Omega_p \int dp\,p^{D-1}\,\epsilon\, f_N(\epsilon).
\ea{calc2}
$k_D$ is the angular part of the momentum space integral, $D$ is the dimension of the phase space, $\Gamma(x)$ is the \textit{Euler-Gamma} function. From Eq.~(\ref{calc2}), it follows that the inverse temperature in each event is proportional to the multiplicity:
\be
\beta_N=\frac{D\,N}{E}.
\ee{calc3}
The multiplicity has Gamma distribution, which is the $\mu = 1$ case of Eq.~(\ref{calc5}) in Ref.~\cite{bib:kno}:
\be
p(N)=A_m\, N^{\alpha -1} e^{-\beta\, N},
\ee{calc4}
with fit parameters $A_m, \alpha, \beta$. In this case, the average momentum distribution is TS:
\be
\frac{\dd\sigma}{\dd^Dp} = \sum p(N)\,N\, f_N(\epsilon) \approx \frac{ \kappa_{D,E}\; }{ \left( 1 + \frac{D}{\beta} x \right)^{\alpha + D + 1}}.
\ee{calc5}
The discrete sum has been approximated by a continuous integral, $x=\epsilon/E$, the energy of the produced hadron scaled by the maximal acquirable energy in a 2-jet event $E=\sqrt{s}/2$, and $\kappa_{D,E}=A_m\,D^D\,\beta^{\,\alpha+D+1}/(E^D\,k_D\,\Gamma(D)\,)$. Note that from Eqs.~(\ref{calc3}) and~(\ref{calc4}) it follows that the inverse temperature, $\beta_N$, is also Gamma-distributed. This can also lead to TS momentum distribution (see Refs.~\cite{bib:tsinppNN11,bib:tsinppNN12}).

In case \textit{\textbf{b}}, the normalised one-particle distribution in events with multiplicity $N$ is
\be
 f_N(\epsilon) = A_{mc}\; (1-x)^{D(N-1)-1},
\ee{calc6}
where $A_{mc} = {D\,N - 1 \choose D\,(N-1)-1}\, D /(\, k_D\, E^D\,)$ follows from the condition
\be
1 = \int d\Omega_p \int dp\,p^{D-1} f_N(\epsilon).
\ee{calc7}
Eq. (\ref{calc6}) follows from the microcanonical momentum-space volume at fixed energy and multiplicity,
\ba
\Omega_{N}(E) &=& \frac{1}{N!}\int \prod d^Dp_i \:\delta\left( E-\sum\epsilon_i\right)\:= \nl
              &=& \frac{k_D^N\,\Gamma(D)}{N!} E^{N\,D-1},
\ea{calc8}
from which the one-particle distribution is obtained as
\be
f_N(\epsilon) \propto \frac{\Omega_{N-1}(E-\epsilon)}{\Omega_N(E)}.
\ee{calc9}
At this point, it is exploited that a shift in the multiplicity ($N\rightarrow N-N_0$) may be made without violating the KNO scaling (see Eq.~(2) in \cite{bib:kno}), and the averaging is done over the multiplicity distribution,
\be
p(N)=A_m\, (N-N_0)^{\alpha -1} e^{-\beta\, (N-N_0)}.
\ee{calc10}
The resulting momentum distribution is a possible microcanonical generalisation of the TS:
\be
\frac{d\sigma}{d^Dp} \propto \frac{1-x}{ \left( 1 - \frac{D}{\beta} \ln(1-x) \right)^{\alpha + D + 1}}.
\ee{calc11}
The discrete sum has again been replaced by a continuous integral, and the factorials in Eq.~(\ref{calc6}) have been approximated by the next-to-leading order Stirling-formula. The shift in the multiplicity has been chosen to be $N_0 = 1+2/D$. This choice of $N_0$ may seem rather \textit{ad hoc}, but fittings shown in Section~\ref{sec:res} (Figs.~\ref{fig:mp1DN3}-\ref{fig:mp1DN3b}) justify it.

A similar formula (Eq.~(\ref{calc11}) without the $1-x$ factor in the numerator) arises in the canonical approach for a system where both the entropy and energy are composed
using non-additive rules~\cite{BiroVan:2011}.

\section{Results}
\label{sec:res}
\begin{figure}
\begin{center}
 \includegraphics[width=8cm,height=8cm]{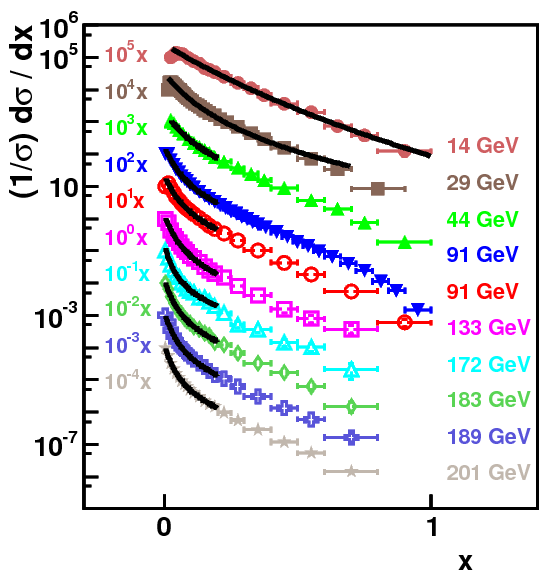}
\end{center}
\caption{ \label{fig:1Dts}
Fragmentation functions measured at various collision energies (data of graphs from top to bottom are published in refs. \cite{bib:tassoX14,bib:tpcX29,bib:tassoX14,bib:delphiX91,bib:opalX91,bib:opalX133,bib:opalX180,bib:opalX180,bib:opalX180,bib:opalX200}) and fitted 1 dimensional TS distributions (Eq.~(\ref{res1}) with $D=1$).
}
\end{figure}

\begin{figure}
\begin{center}
 \includegraphics[width=8cm,height=8cm]{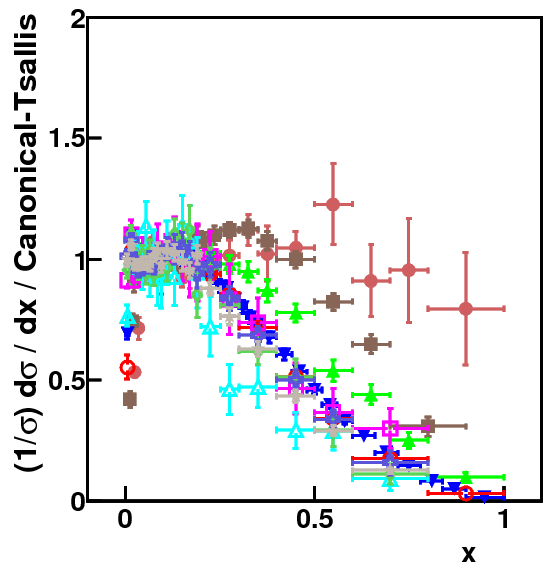}
\end{center}
\caption{ \label{fig:1DtsDperT}
Ratios of measured fragmentation functions and fitted 1 dimensional TS distributions (Eq.~(\ref{res1}) with $D=1$) at various collision energies (data of graphs are published in refs. \cite{bib:tassoX14}-\cite{bib:opalX200}).}
\end{figure}

\begin{figure}
\begin{center}
 \includegraphics[width=8cm,height=8cm]{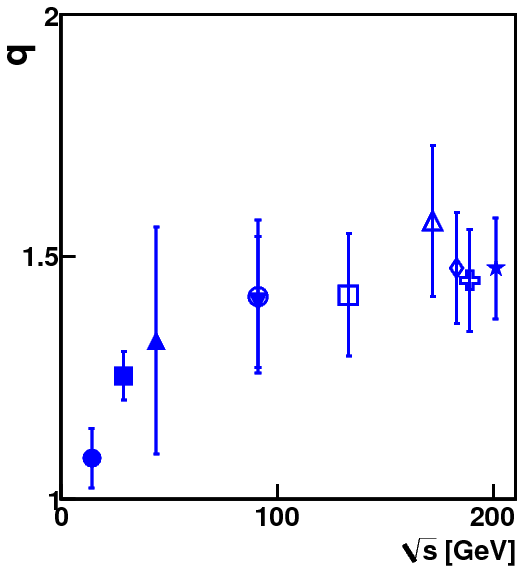}
\end{center}
\caption{ \label{fig:1DtsQ}
Fitted values of the $q$ parameter in Eq.(\ref{res1}) with $D=1$ to measured fragmentation functions shown in Fig.~\ref{fig:1Dts}.}
\end{figure}

\begin{figure}[hp]
\begin{center}
 \includegraphics[width=8cm,height=8cm]{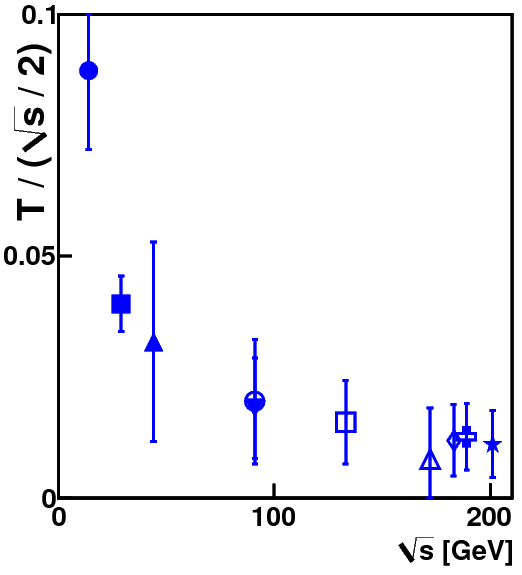}
\end{center}
\caption{ \label{fig:1DtsT}
Fitted values of the $T$ parameter in Eq.~(\ref{res1}) with $D=1$ to measured fragmentation functions shown in Fig.~\ref{fig:1Dts}.}
\end{figure}

When fitting data on fragmentation, we used the parametrisation of the TS distribution:
\be
\frac{1}{\sigma}\frac{d\sigma}{dx} = A x^{D-1} \left( 1 + \frac{q-1}{T/(\sqrt{s}/2)}x\right)^{-1/(q-1)},
\ee{res1}
whith the parameters $q = 1 + 1/(\alpha+D+1)$ and $T = (\sqrt{s}/2)\,\beta/(\,D\,(\alpha+D+1)\,)$. Fittings of Eq.~(\ref{res1}) to measured fragmentation functions (Fig.~\ref{fig:1Dts}) and ratios of data and fits (Fig.~\ref{fig:1DtsDperT}) show that the one-dimensional TS reproduces data for about $0.01\lessapprox x \lessapprox 0.2$. The increase of $q$ (Fig.~\ref{fig:1DtsQ}) and decrease of $T$ (Fig.~\ref{fig:1DtsT}) parameters with $\sqrt{s}$ is similar to what was found in \cite{bib:tsinppNN9b}, and differs from $pp$ results~\cite{bib:tsinppNN9}, where $T$ was found to be independent of $\sqrt{s}$ (though the analysis was done only for $\sqrt{s}>$ 200 GeV).

\begin{figure}[hp]
\begin{center}
 \includegraphics[width=8cm,height=8cm]{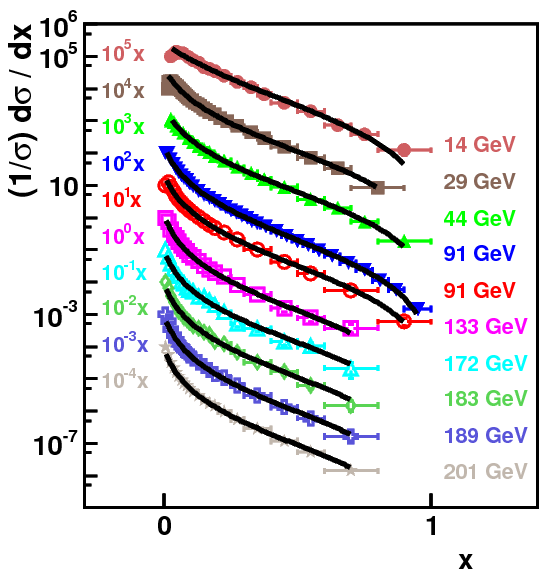}
\end{center}
\caption{ \label{fig:1Dmts}
Fragmentation functions measured at various collision energies (data of graphs from top to bottom are published in refs. \cite{bib:tassoX14,bib:tpcX29,bib:tassoX14,bib:delphiX91,bib:opalX91,bib:opalX133,bib:opalX180,bib:opalX180,bib:opalX180,bib:opalX200}) and fitted 1 dimensional MTS distributions (Eq.~(\ref{res2}) with $D=1$).}
\end{figure}

\begin{figure}[hp]
\begin{center}
 \includegraphics[width=8cm,height=8cm]{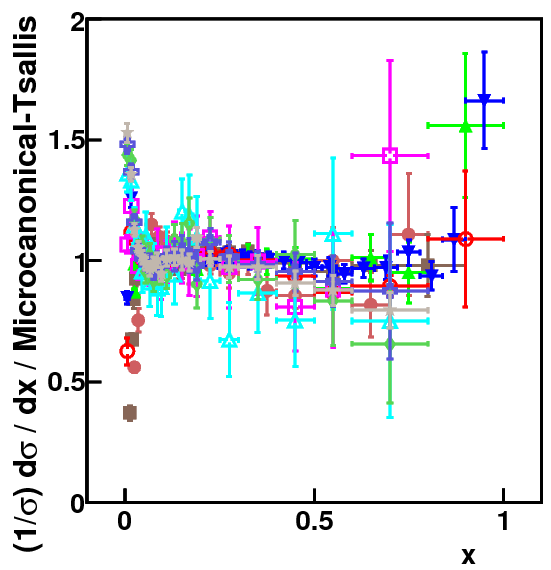}
\end{center}
\caption{ \label{fig:1DmtsDperT}
Ratios of measured fragmentation functions and fitted 1 dimensional MTS distributions (Eq.~(\ref{res2}) with $D=1$) at various collision energies (data of graphs are published in refs. \cite{bib:tassoX14}-\cite{bib:opalX200}).}
\end{figure}

\begin{figure}[hp]
\begin{center}
 \includegraphics[width=8cm,height=8cm]{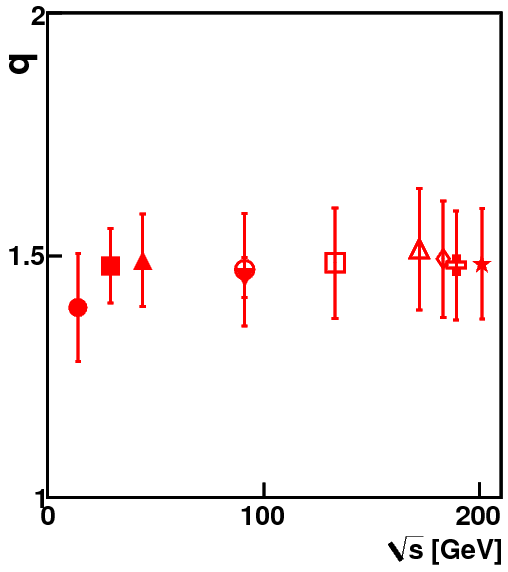}
\end{center}
\caption{ \label{fig:1DmtsQ}
Fitted values of the $q$ parameter in Eq.~(\ref{res2}) with $D=1$ to measured fragmentation functions shown in Fig.~\ref{fig:1Dmts}.}
\end{figure}

\begin{figure}[hp]
\begin{center}
 \includegraphics[width=8cm,height=8cm]{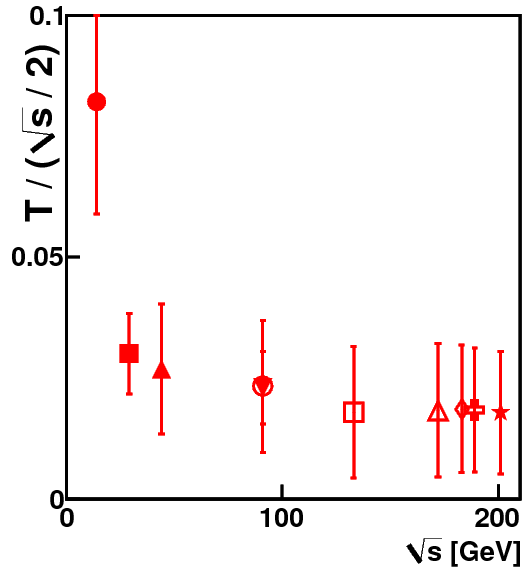}
\end{center}
\caption{ \label{fig:1DmtsT}
Fitted values of the $T$ parameter in Eq.~(\ref{res2}) with $D=1$ to measured fragmentation functions shown in Fig.~\ref{fig:1Dmts}.}
\end{figure}

Rewriting the microcanonical TS (Eq.~(\ref{calc11})) in terms of the parameters, $q$ and $T$, we arrive at
\be
\frac{1}{\sigma}\frac{d\sigma}{dx} = \frac{A\, x^{\,D-1} \left( 1-x \right) } {\left( 1 - \frac{q-1}{T/(\sqrt{s}/2)}\ln(1-x)\right)^{1/(q-1)}} \;.
\ee{res2}
Fittings of this microcanonical generalisation of the TS in one-dimension to fragmentation functions (Fig.~\ref{fig:1Dmts}) as well as ratios of data and fits (Fig.~\ref{fig:1DmtsDperT}), show that Eq.~(\ref{res2}) gives a good fit in a wide range, $0.01\lessapprox x \lessapprox 1$. The increase of $q$ with $\sqrt{s}$ disappears (Fig.~\ref{fig:1DmtsQ}), and the decrease of $T$ with $\sqrt{s}$ reduces significantly (Fig.~\ref{fig:1DmtsT}).

\begin{figure}[hp]
\begin{center}
 \includegraphics[width=8cm,height=8cm]{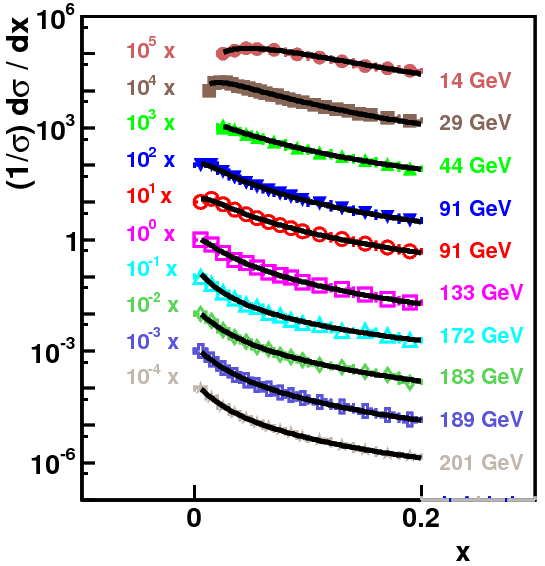}
\end{center}
\caption{ \label{fig:3to1Dts}
Fragmentation functions measured at various collision energies (data of graphs from top to bottom are published in refs. \cite{bib:tassoX14,bib:tpcX29,bib:tassoX14,bib:delphiX91,bib:opalX91,bib:opalX133,bib:opalX180,bib:opalX180,bib:opalX180,bib:opalX200}) and fitted TS distributions (Eq.~(\ref{res1}) with dimension $D$, decreasing like $D\sim1/\sqrt{s}$ from $D=3$ down to $D=1$).}
\end{figure}

\begin{figure}[hp]
\begin{center}
 \includegraphics[width=8cm,height=8cm]{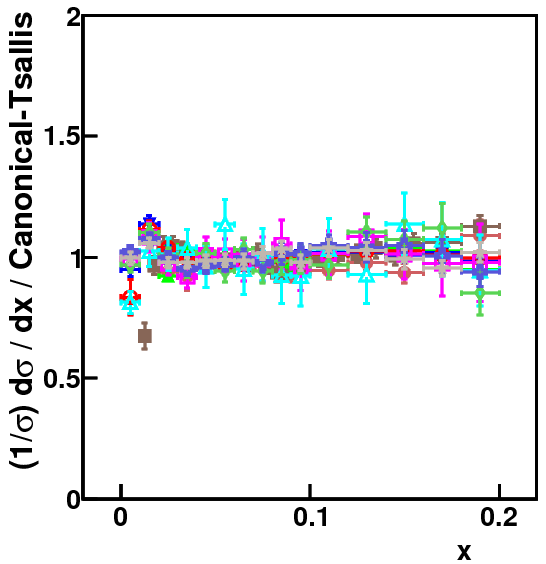}
\end{center}
\caption{\label{fig:3to1DtsDperT}
Ratios of measured fragmentation functions and fitted TS distributions (Eq.~(\ref{res1}) with dimension $D$, decreasing like $D\sim1/\sqrt{s}$ from $D=3$ down to $D=1$) at various collision energies (data of graphs are published in refs. \cite{bib:tassoX14}-\cite{bib:opalX200}).}
\end{figure}

Both the canonical and the microcanonical TS in one dimension fail to describe data for very low $x$ ($x \lessapprox 0.01$). Introducing a scaling in the dimension pa\-ra\-me\-ter, $D\sim 1/\sqrt{s}$ solves this problem, as can be seen in Figs.~\ref{fig:3to1Dts} and~\ref{fig:3to1DtsDperT} showing fits. The interpretation of this is not straightforward, though. It should not mean that hadron production in a jet at $\sqrt{s}=200$ GeV is a one-dimensional process, while at $\sqrt{s}=14$ GeV it is a three-dimensional one. Let us consider that transverse momenta ($p_T$) of hadrons in a jet (transverse with respect to the jet axis) are smaller than $p_{T0}=2$ GeV/c. Then the ratio of the maximal transverse hadron momentum and the maximal longitudinal hadron momentum ($p_{L0}=\sqrt{s}/2$ considering 2-jet events) grows from $p_{T0}/p_{L0} \approx 0.02$ up to $p_{T0}/p_{L0} \approx 0.3$ as $\sqrt{s}$ decreases from $\sqrt{s}=200$ GeV down to $\sqrt{s}=14$ GeV. Hadrons with momenta $p<p_{T0}$ (which means $x<p_{T0}/(\sqrt{s}/2)$) may be produced isotropically in the hemi-sphere of the jet, while hadrons with momenta $p>>p_{T0}$ are predestined to go approximately parallelly to the jet axis. So, the effective dimension of the phase space depends on $p/p_{T0}$, rather than on $x$ or on $\sqrt{s}$. Thus this causes scaling violation of the fragmentation functions. In non-perturbative models~\cite{bib:npert1}--\cite{bib:npert3} a similar, $(m_T/E_{jet})^a$ (with $a\sim1-2$), energy dependence of spectra ($m_T$ being the transverse energy of the hadron, of order 0.5 GeV) was reported as well.

\begin{figure}[hp]
\begin{center}
 \includegraphics[width=8cm,height=8cm]{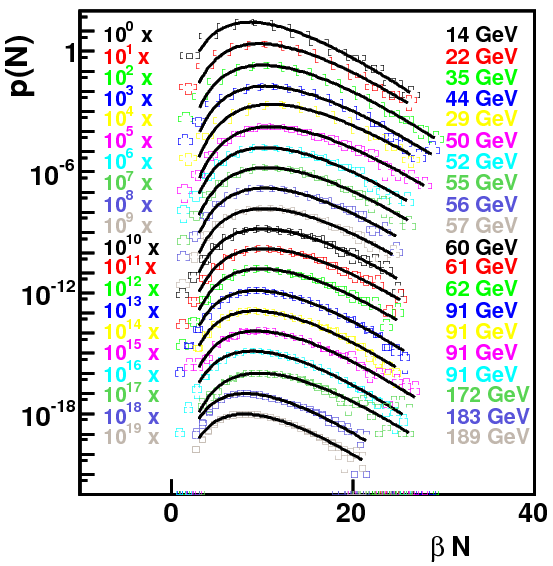}
\end{center}
\caption{ \label{fig:mpN0}
Multiplicity distributions measured at various collision energies (data of graphs from top to bottom are published in refs. \cite{bib:tassoN14to44,bib:tassoN14to44,bib:tassoN14to44,bib:tassoN14to44,
bib:hrsN29,
bib:amyN50to62,bib:amyN50to62,bib:amyN50to62,bib:amyN50to62,bib:amyN50to62,bib:amyN50to62,bib:amyN50to62,bib:amyN50to62,
bib:delphiN91,
bib:L3N91,
bib:alephN91,
bib:opalN91,
bib:opalX180,bib:opalX180,bib:opalX180}) and fitted Gamma-distributions (Eq.~(\ref{calc10}) with $N_0=0$).}
\end{figure}

\begin{figure}[hp]
\begin{center}
 \includegraphics[width=8cm,height=8cm]{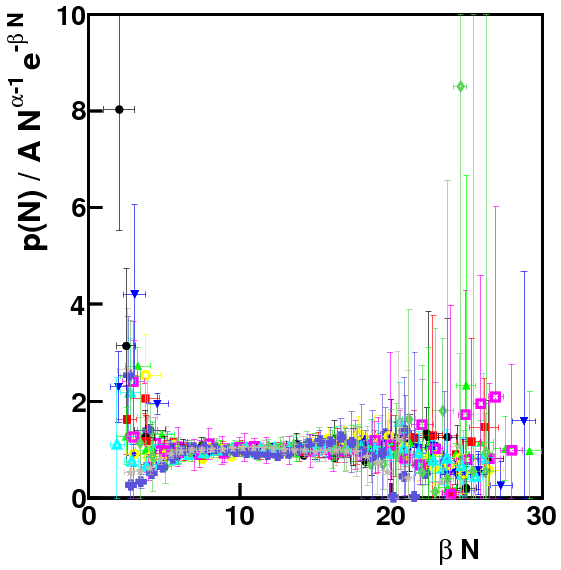}
\end{center}
\caption{ \label{fig:mpN0DperT}
Ratios of measured multiplicity distributions and fitted Gamma-distributions (Eq.~(\ref{calc10}) with $N_0=0$) measured at various collision energies (data of graphs are published in refs. \cite{bib:opalX180} and \cite{bib:tassoN14to44}-\cite{bib:opalN91}).}
\end{figure}

\begin{figure}[hp]
\begin{center}
 \includegraphics[width=8cm,height=8cm]{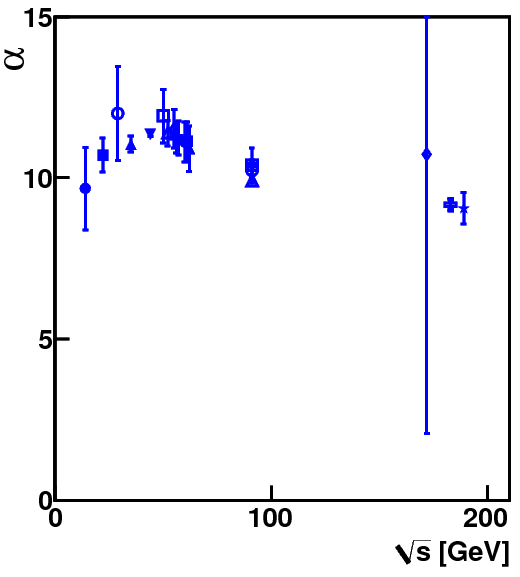}
\end{center}
\caption{ \label{fig:mpN0a}
Fitted values of the $\alpha$ parameter in Eq.~(\ref{calc10}) with $N_0=0$ to the measured multiplicity distributions shown in Fig.~\ref{fig:mpN0}.}
\end{figure}

\begin{figure}[hp]
\begin{center}
 \includegraphics[width=8cm,height=8cm]{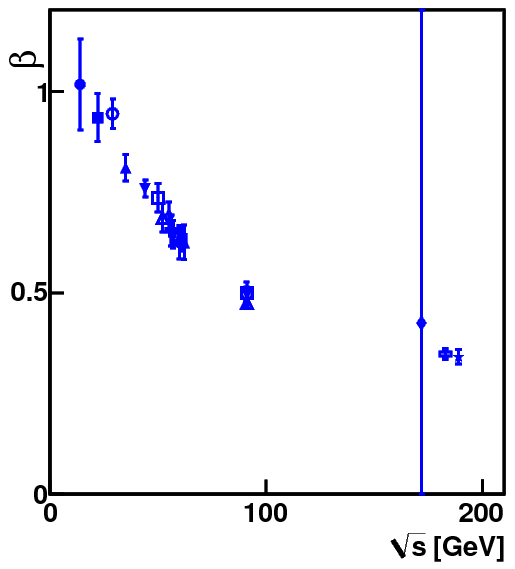}
\end{center}
\caption{ \label{fig:mpN0b}
Fitted values of the $\beta$ parameter in Eq.~(\ref{calc10}) with $N_0=0$ to the measured multiplicity distributions shown in Fig.~\ref{fig:mpN0}.}
\end{figure}

\begin{figure}[hp]
\begin{center}
 \includegraphics[width=8cm,height=8cm]{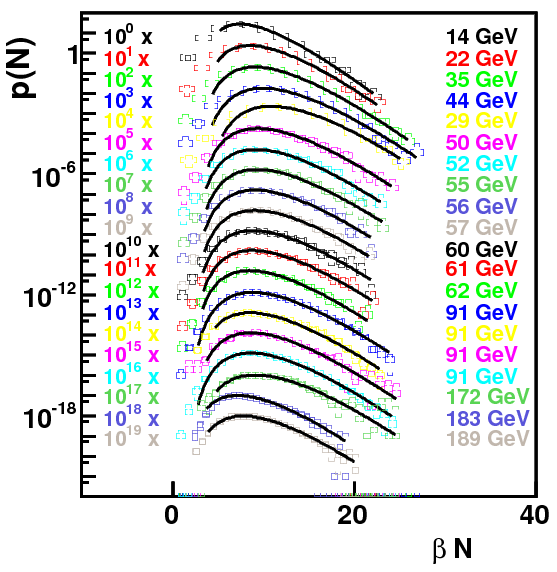}
\end{center}
\caption{ \label{fig:mp1DN3}
Multiplicity distributions measured at various collision energies (data of graphs from top to bottom are published in refs. \cite{bib:tassoN14to44,bib:tassoN14to44,bib:tassoN14to44,bib:tassoN14to44,
bib:hrsN29,
bib:amyN50to62,bib:amyN50to62,bib:amyN50to62,bib:amyN50to62,bib:amyN50to62,bib:amyN50to62,bib:amyN50to62,bib:amyN50to62,
bib:delphiN91,
bib:L3N91,
bib:alephN91,
bib:opalN91,
bib:opalX180,bib:opalX180,bib:opalX180}) and fitted shifted Gamma-distributions (Eq.~(\ref{calc10}) with $N_0=3$).}
\end{figure}

\begin{figure}[hp]
\begin{center}
 \includegraphics[width=8cm,height=8cm]{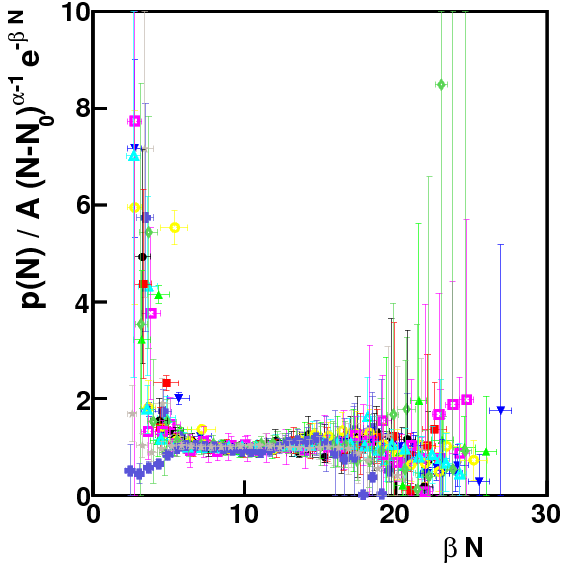}
\end{center}
\caption{ \label{fig:mp1DN3DperT}
Ratios of measured multiplicity distributions and fitted shifted Gamma-distributions (Eq.~(\ref{calc10}) with $N_0=3$) measured at various collision energies (data of graphs are published in refs. \cite{bib:opalX180} and \cite{bib:tassoN14to44}-\cite{bib:opalN91}).}
\end{figure}

\begin{figure}[hp]
\begin{center}
 \includegraphics[width=8cm,height=8cm]{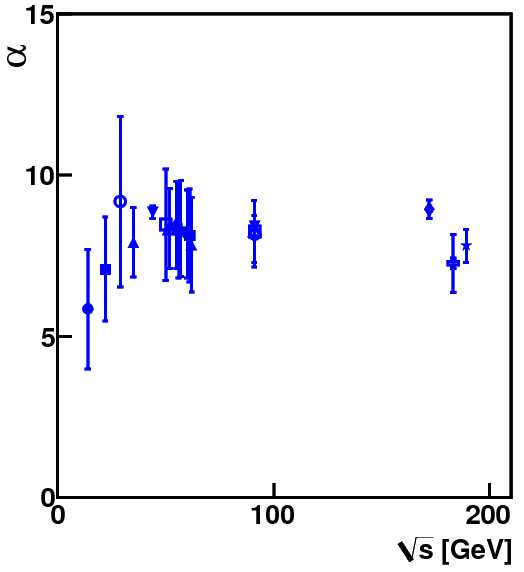}
\end{center}
\caption{ \label{fig:mp1DN3a}
Fitted values of the $\alpha$ parameter in Eq.~(\ref{calc10}) with $N_0=3$ to the measured multiplicity distributions shown in Fig.~\ref{fig:mp1DN3}.}
\end{figure}

\begin{figure}[hp]
\begin{center}
 \includegraphics[width=8cm,height=8cm]{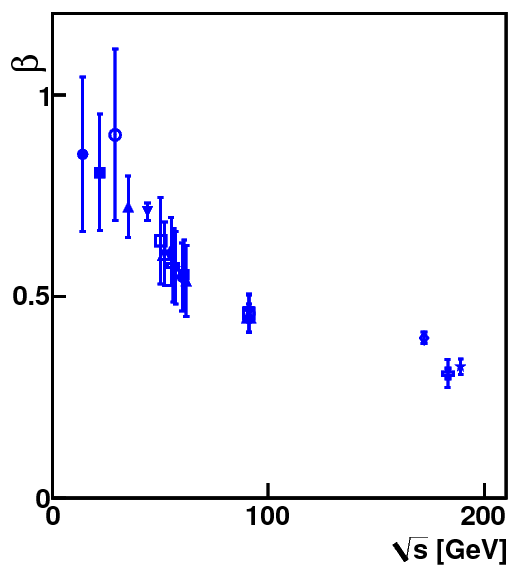}
\end{center}
\caption{ \label{fig:mp1DN3b}
Fitted values of the $\beta$ parameter in Eq.~(\ref{calc10}) with $N_0=3$ to the measured multiplicity distributions shown in Fig.~\ref{fig:mp1DN3}.}
\end{figure}

In Section~\ref{sec:calc}, it has been shown that a Gamma-distribution of the shifted multiplicity $N-N_0$ can result in a TS or microcanonical TS shaped spectrum. Fits of Eq.~(\ref{calc4}) (shown in Figs.~\ref{fig:mpN0}-\ref{fig:mpN0b}) and of Eq.~(\ref{calc10}) (shown in Figs.~\ref{fig:mp1DN3}-\ref{fig:mp1DN3b}) to data on multiplicity distributions measured at various collision energies, show that both assumptions, $N_0=0$ and $N_0=1+2/D$ (with $D=1$) give acceptable agreement with data. However, the predicted power of the spectrum from Eq.~(\ref{calc11}), scattering around 12 and 10 (Figs.~\ref{fig:mpN0a} and~\ref{fig:mp1DN3a}), is much higher than what is seen from fits to data on fragmentation functions. The fitted power, $1/(q-1)$ in Eq.~(\ref{res1}) and in Eq.~(\ref{res2}), scatters around 2 and 2-4 respectively (Fig.~\ref{fig:1DtsQ} and Fig.~\ref{fig:1DmtsQ}).


\section{Conclusions}
\label{sec:con}
In this paper, we compared inclusive data on $e^+e^-\rightarrow h^{\pm}+X$ reactions with the models on statistical hadron production described in Section~\ref{sec:calc}. In these models, we considered 2-jet events, neglected the masses of quarks and hadrons. Furthermore we assumed that hadrons in a single jet are produced according to a statistical distribution. We used two types of distributions: the Boltzmann-Gibbs, and the microcanonical distribution in one dimension, for the reasons discussed in Section~\ref{sec:calc}.

We found that event-by-event fluctuations of the multiplicity can be described well by the Gamma-distribution, and that such fluctuations can result in an average fragmentation function of the form of the Tsallis-like, or a possible microcanonical generalisation of the Tsallis distribution. In Section~\ref{sec:res} we have shown that Tsallis distribution reproduces data on fragmentation for $0.01 \lessapprox x \lessapprox 0.2$, while the microcanonical Tsallis for $0.01 \lessapprox x \lessapprox 1$. Below $x\approx 0.01$, the one-dimensional models tested, deviate from measurements, especially for $\sqrt{s}\leq M_Z$.

Introduction of an effective phase-space dimension, scaling like $D\sim1/\sqrt{s}$ seemingly solves the problem. However, the underlying physics, as argued in Section~\ref{sec:res}, supports the emergence of a scale-breaking term in the fragmentation function of the form, $(m_T/E_{jet})^a$, as  reported in Refs.~\cite{bib:npert1}-\cite{bib:npert3}, rather than a scaling of the dimensionality of the phase space. Furthermore, when considering low-$x$ hadron production, yields from 3-jet events are not negligble, especially for $\sqrt{s}<M_Z$.

Anyway, the decisive argument for or against the models introduced above would be the measurement of momentum distribution of hadrons in single jets for fixed multiplicity.

\section*{Acknowledgement}
\label{sec:ack}
This work was supported by the Hungarian OTKA grants PD73596, K68108 and the E\"otv\"os University. One of the authors (GGB) thanks the J\'anos Bolyai Research Scolarship of the Hungarian Academy of Sciences.








\end{document}